\documentclass[aps,pre,twocolumn,groupedaddress]{revtex4-1} 
 
\usepackage{graphicx}
\usepackage{dcolumn}
\usepackage{amsmath}
\usepackage{amssymb}
\usepackage{bm}

\begin{document}


\title{Membrane structure formation induced by two types of banana-shaped proteins}

\author{Hiroshi Noguchi}
\email[]{noguchi@issp.u-tokyo.ac.jp}
\affiliation{Institute for Solid State Physics, University of Tokyo,
 Kashiwa, Chiba 277-8581, Japan}
\author{Jean-Baptiste Fournier}
\email[]{jean-baptiste.fournier@univ-paris-diderot.fr}
\affiliation{Laboratoire Mati\`ere et Syst\`emes  Complexes (MSC), UMR 7057 CNRS, Universit\'e Paris Diderot, F-75205, Paris, France}
\begin{abstract}
The assembly of banana-shaped rodlike proteins on membranes, and the associated membrane shape transformations, 
are investigated by analytical theory and coarse-grained simulations.
The membrane-mediated interactions between two banana-shaped inclusions 
are derived theoretically using a point-like formalism based on fixed anisotropic curvatures, both for zero surface tension and for finite surface tension. On a larger scale, the interactions between assemblies of such rodlike inclusions are determined analytically.
Meshless membrane simulations are performed in the presence of a large number of inclusions of two types, corresponding to curved rods of opposite curvatures, both for flat membranes and vesicles.
Rods of the same type aggregate into linear assemblies perpendicular to the rod axis, leading to membrane tubulation.
However, rods of the other type, those of opposite curvature, are attracted to the lateral sides of these assemblies, and stabilize a straight bump structure that prevents tubulation.
When the two types of rods have almost opposite curvatures,
the bumps attract one another, forming a stripe structure.
Positive surface tension is found to stabilize the stripe formation.
The simulation results agree well with the theoretical predictions provided the point-like curvatures of the model are scaled-down to account for the effective flexibility of the simulated rods.
\end{abstract}

\maketitle

\section{Introduction}

In living cells, membrane shape transformation plays a key role in biological functions such as endo/exocytosis and vesicle transports.
Cell organelles have specific shapes depending on their functions.
Various types of proteins participate in the regulation of these dynamic and static membrane shapes \cite{mcma05,shib09,drin10,baum11,joha14,mcma15}.
These proteins mainly control local membrane shapes in two ways: hydrophobic insertions (wedging) and scaffolding.
In the former mechanism, a part of the protein, such as an amphipathic $\alpha$-helix, is inserted into the lipid bilayer membrane.
In the latter mechanism, the protein domain has a strong affinity for the lipid polar head groups and adsorbs onto the lipid membrane.
A BAR (Bin/Amphiphysin/Rvs) domain, which consists of a banana-shaped dimer,
mainly bends the membrane along the domain axis via scaffolding~\cite{itoh06,masu10,zhao11,mim12a,simu15a}.
Some of the BAR superfamily proteins, such as N-BAR proteins, also have hydrophobic insertions.
Experimentally, the membrane tubulation and curvature-sensing by various types of BAR superfamily proteins have been observed~\cite{itoh06,masu10,zhao11,mim12a,simu15a,pete04,matt07,fros08,wang09,zhu12,tana13,shi15,prev15,isas15,adam15,simu16}.

Objects with rotational symmetry, such as spherical colloids or conical integral proteins inserted perpendicularly to the membrane,  generate an isotropic membrane curvature. Conversely, BAR domains which are banana-shaped generate an anisotropic curvature~\cite{lipo13,reyn07,auth09,sari12} (amphipathic $\alpha$-helices can also yield an anisotropic curvature~\cite{gome16}). Theoretical models have shown that membrane inclusions, such as adsorbed or embedded proteins, or colloids, undergo long-range interactions that are mediated by the curvature elasticity of the membrane~\cite{Goulian93EPL}. It was also shown that anisotropic inclusions experience interactions of longer range than isotropic ones and are able to produce complex aggregates~\cite{Park96JPhysI,domm99}. 
To simplify the theoretical calculations, membrane inclusions are usually modelled as non-deformable objects with a fixed curved shape~\cite{domm99,domm02,four03,schw15}. Owing to their small sizes  it is often convenient to treat them as point-like objects~\cite{domm99,domm02,Yolcu14ACIS}.
Although the existence of membrane--mediated interactions has been verified experimentally in the case of isotropic inclusions~\cite{vanderWel:2016du}, there is no direct experimental evidence yet in the case of anisotropic proteins, despite their biological importance.

Numerical simulations are therefore essential in this context.
Atomic and coarse-grained molecular simulations \cite{bloo06,arkh08,yu13,simu13,simu15} 
have been employed to investigate molecular-scale interactions between BAR proteins and lipids. 
The scaffold formation~\cite{yu13} and linear assembly~\cite{simu13} of BAR domains have been demonstrated.
To investigate large-scale membrane deformations,
a dynamically triangulated membrane model \cite{rama12,rama13} 
and meshless membrane models \cite{nogu14,nogu15b,nogu16,nogu16a,ayto09} have been employed;
consequently, various (meta)stable vesicle shapes~\cite{rama12,rama13,nogu14,nogu15b}
and the tubule formation dynamics \cite{nogu16} have been reported.
However, the relation with the large-scale picture of the theory of point-like anisotropic inclusions~\cite{domm99,domm02,four03} is not well investigated.
Here, we compare the meshless membrane simulations with this theory.

In living cells, more than one type of BAR and other proteins cooperatively work to regulate  membrane shape.
However, in most of the previous reports,
rods of a single type are considered in theories and simulations.
To our knowledge, only two studies have been reported for the interactions of two types of proteins.
The mixture of inclusions with isotropic and anisotropic curvatures was shown to produce the self-assembly of a neckless of anisotropic inclusions around a domain consisting of a lattice of isotropic inclusions, mimicking the assembly of dynamin proteins around a scaffolded bud~\cite{four03}.
The phase segregation of rods with different positive spontaneous curvatures was reported in Ref.~\citenum{rama13}.

In the present study, we investigate the membrane-mediated interactions between two types of protein rods with opposite spontaneous curvatures, 
theoretically and numerically.
This corresponds to the situations in which oppositely curved proteins, such as I-BAR and other BAR proteins, are adsorbed on the same leaflet of a bilayer membrane,
or alternatively, two types of positively curved proteins are adsorbed on opposite leaflets.

In our theoretical analysis we use a multi-scale approach. First we consider the interaction between two rods of possibly different curvatures. We assume that the rods are separated by a distance larger than their size so that we can treat them as point-like anisotropic inclusions using the Green function formalism of Refs.~\citenum{domm02,four03}. Contrary to previous models~\cite{domm99,domm02,four03} we assume here, in order to model banana-shaped BAR domains, that the inclusions fix a curvature in the rod's direction but do not impose any curvature in the orthogonal direction. We also consider the interaction under nonzero surface tension, contrary to what was done in previous works~\cite{domm99,domm02,four03}. Our results show that rods of same curvature self-assemble into long straight structures that attract rods of opposite curvature on their sides. We then change the scale of our analysis and study the interaction between these macroscopic straight structures. We find that their interaction depends crucially on membrane tension.

In our simulations an implicit-solvent meshless membrane model~\cite{nogu09,nogu06,shib11,nogu14,nogu15b,nogu16,nogu16a}
is used to represent a fluid membrane.
Banana-shaped proteins, assumed to be strongly adsorbed onto the membrane, are modelled together with the membrane region below them as linear strings of particles with a bending stiffness and a preferred curvature.
In order to investigate the membrane-mediated interactions,
no direct attractive interaction is considered between the rods. We investigate the interaction and structures produced by a mixture of a large number of rods of two types. We find that the results of our simulations agree very well with the theoretical predictions provided the point-like curvatures are scaled-down to account for the rods flexibility.

In Sec.~\ref{sec:theory}, our multi-scale theoretical analysis of the interactions between curved rods adsorbed onto a membrane with bending rigidity and tension is presented.
In Sec.~\ref{sec:method}, the simulation model and method are described. 
In Sec.~\ref{sec:2rods},  the simulation and theoretical results  are compared for the interactions of two rods.
In Secs.~\ref{sec:flat} and~\ref{sec:ves}, the assembly of protein rods in flat membranes and vesicles are presented.
The summary and discussion are given in Sec.~\ref{sec:sum}.

\section{Theory}\label{sec:theory}

\subsection{Interactions between two curved rods}\label{sec:Tpoint}

We consider BAR-like membrane inclusions shaped as rods of length $r_{\rm {rod}}$ that are
curved in a plane perpendicular to the plane of the membrane. 
In order to compute the interaction between such inclusions, at separations well larger than $r_\mathrm{rod}$, we model them as point-like inclusions that impose some membrane curvature $C$ along the rod direction 
and no constraint along the orthogonal direction.
To derive the interaction between two such rods we use the Green function formalism of Refs.~\citenum{domm02,four03}. We first discuss the case of a tensionless membrane, then we take into account membrane tension. In this model, in order to make the calculations tractable, we neglect two aspects: the finite length of the rods and their effective flexibility. However, as we shall see, these assumptions are not critical.  

\begin{figure}
\centerline{\includegraphics[width=.6\columnwidth]{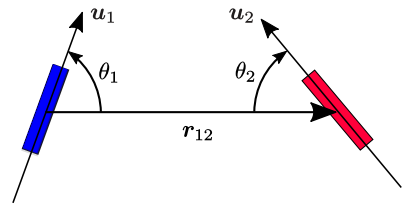}}
\caption{Geometrical parameters for  two rodlike inclusions.}
\label{fig:geom}
\end{figure}

\subsubsection{Tensionless membrane }\label{sec:theory_tensionless}

We first consider two such rods adsorbed onto a membrane with vanishing tension ($\gamma=0$). Let $\bm{r}_{12}$ be the vector going from the center of rod~1 to the center of rod~2, and $\bm{r}_{21}=-\bm{r}_{12}$. We assume that the rods impose curvatures $C_\mathrm{r1}$ and $C_\mathrm{r2}$ along the directions $\bm{u}_1$ and $\bm{u}_2$, at angles $\theta_1$ and $\theta_2$ with respect $\bm{r}_{12}$ and $\bm{r}_{21}$, respectively (Fig.~\ref{fig:geom}). To preserve the symmetry between the two inclusions, we orient $\theta_1$ counterclockwise and $\theta_2$ clockwise. We call $R=|\bm{r}_{12}|$ the distance between the rods.

\begin{figure*}
\centerline{\includegraphics[width=1.6\columnwidth]{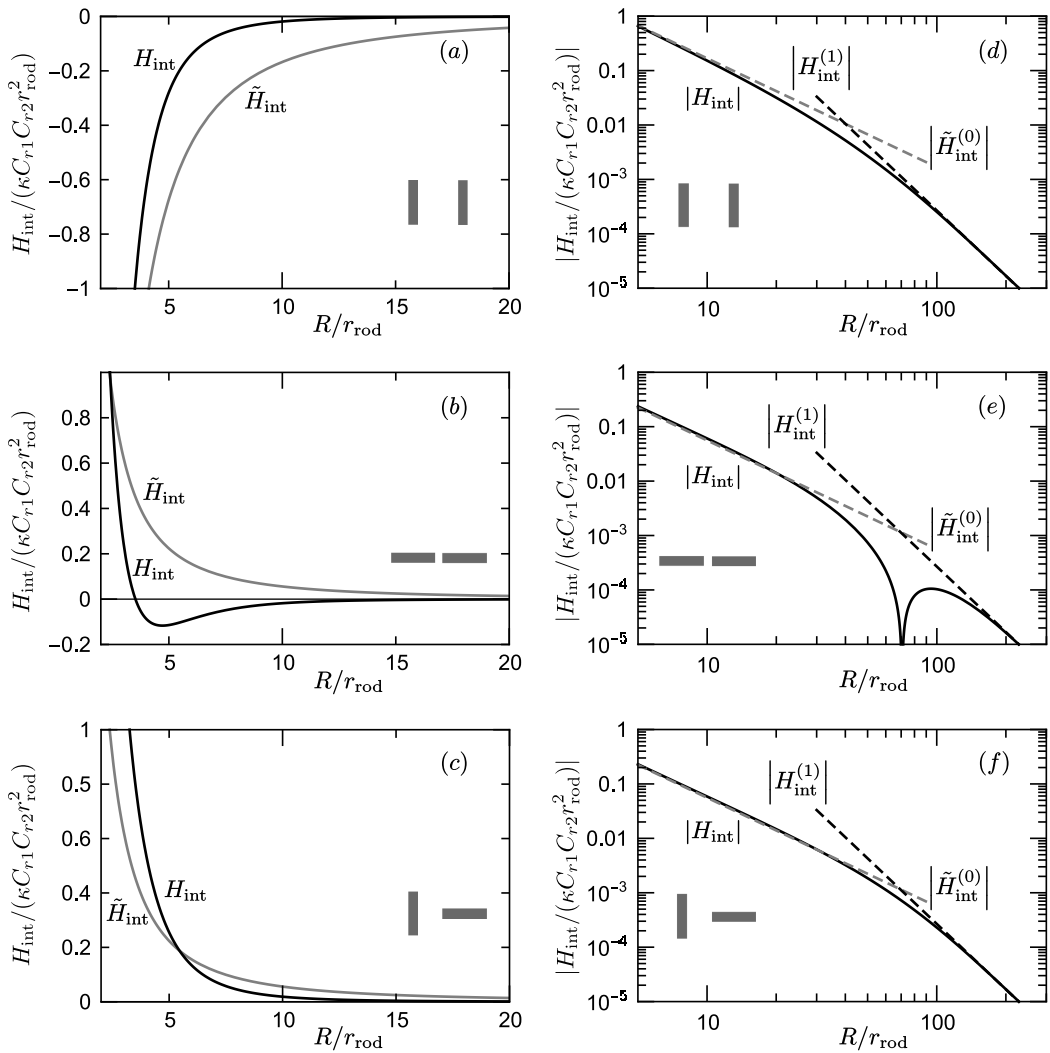}}
\caption{Normalized interaction energy between two rods at vanishing tension, $\tilde H_\mathrm{int}$ (gray curves, plain and dashed), and at non-vanishing tension, $H_\mathrm{int}$ (black curves, plain and dashed). Left column: comparison between the interaction $\tilde H_\mathrm{int}$ at zero tension (gray) and the interaction $H_\mathrm{int}$ at a strong tension corresponding to $\xi/r_\mathrm{rod}=1$ (black). Right column: Logarithmic plot of the interaction at weak tension, for $\xi/r_\mathrm{rod}=20$, showing the crossover at $R\approx3\xi$ from the zero-tension behavior $\sim\!\tilde H_\mathrm{int}^{(0)}$ to the nonzero-tension behavior $\sim\!H_\mathrm{int}^{(1)}$. Top: side-to-side orientation of the rods, i.e., $\theta_1=\theta_2=\pi/2$. Middle: tip-to-tip orientation of the rods, i.e., $\theta_1=\theta_2=0$. Bottom: tip-to-side orientation of the rods, i.e., $\theta_1=0$ and $\theta_2=\pi/2$.}
\label{fig:th-id}
\end{figure*}

We consider the limit of small membrane deformations.
Minimizing the Helfrich bending energy of the membrane~\cite{Helfrich73}
with the curvature constraints yields an interaction which is a binary quadratic form in $C_\mathrm{r1}$ and $C_\mathrm{r2}$, as shown in Appendix~\ref{sec:apthe}. We find that unless $|C_\mathrm{r1}|\ll|C_\mathrm{r2}|$, or the opposite, the interaction is well approximated by the sole term proportional to $C_\mathrm{r1}C_\mathrm{r2}$ up to distances comparable to $r_\mathrm{rod}$. Accordingly, in the following we neglect the contributions proportional to $C_\mathrm{r1}^2$ and $C_\mathrm{r2}^2$. The curvature-mediated interaction energy between two rods in the absence of membrane tension, $\tilde H_\mathrm{int}(R)$, is thus obtained as
\begin{align}
\tilde H_\text{int}(R)\simeq\tilde H_\text{int}^{(0)}(R)= &
\frac{16 \pi r_\mathrm{rod}^4}{9R^2}\kappa C_{\rm r1} C_{\rm r2} \big[\cos(2\theta_1)+\cos(2\theta_2) \nonumber \\ 
& -\cos(2\theta_1-2\theta_2)\big]. \label{eq:Hint}
\end{align}
Here $\kappa$ is the bending rigidity of the membrane. We have exhibited the leading order term $\propto\!1/R^2$, which describes the interaction quantitatively well for $R$ larger than a few times $r_\mathrm{rod}$. Note that this interaction is of longer range than the $\propto\!1/R^4$ fluctuation-induced Casimir interaction between straight rods and it has a different angular dependence~\cite{Golestanian96PRE,Bitbol11EPL}.
Graphs of this interaction are represented in Fig.~\ref{fig:th-id}a, b and c for various orientations and curvatures of the rods.

When one of the rods is parallel to the separation vector to the other rod ($\theta_1=0$ or $\theta_2=0$),
 eqn~(\ref{eq:Hint}) gives
\begin{align} \label{eq:H0}
\tilde H_\text{int}^{(0)}(R)= \frac{16\pi r_\mathrm{rod}^4}{9R^2}\kappa C_{\rm r1}C_{\rm r2},
\end{align}
which is independent of the orientation of the other rod.

In contrast, when the axes of both rods are perpendicular to the separation vector, i.e., $\theta_1=\theta_2=\pi/2$,
$\tilde H_{\text{int}}^{(0)}$ has the opposite sign and an amplitude three times larger:
\begin{align} \label{eq:H05pi}
\tilde H_\text{int}^{(0)}(R)= -\frac{16\pi r_\mathrm{rod}^4}{3R^2}\kappa C_{\rm r1}C_{\rm r2}.
\end{align}

Hence, when two rods are identical, i.e., $C_{\rm r1}=C_{\rm r2}$,
they have a strong attractive interaction at $\theta_1=\theta_2=\pi/2$,
while they have a weak repulsive interaction at $\theta_1=0$ or $\theta_2=0$.
Obviously, when they are perpendicular to their separation vector they will attract up to contact since they match exactly and produce less deformation when they superimpose.

When two rods have curvatures of opposite sign, i.e.,  $C_{\rm r1}C_{\rm r2}<0$,
the interactions are opposite.
For $\theta_1=0$ or $\theta_2=0$, the rods have a weak attractive interaction. 
Rods of opposite curvatures in a tip-to-tip configuration produce no large-scale curvature deformation when they get close, 
therefore they will attract up to contact. This will likely be true also for rods not exactly opposite in curvature.

Thus, for neighboring rods at equilibrium, we expect identical rods to align preferentially side-to-side and rods of opposite curvature to align tip-to-tip.

\subsubsection{Membrane under tension }\label{sec:theory_tension}

The derivation of the interaction in the presence of membrane tension, $H_\mathrm{int}(R)$, is described in Appendix~\ref{sec:apthe}, Sec.~\ref{tensioncase}. For the same reasons as previously, we retain in the interaction only the terms proportional to $C_{\rm r1}C_{\rm r2}$. In the regime of strong tensions, i.e., for $\xi=(\kappa/\gamma)^{1/2}\approx r_\mathrm{rod}$, Fig.~\ref{fig:th-id}a, b, and c, show that the interaction $H_\mathrm{int}$ displays several similitudes and differences with the tensionless case:
\begin{enumerate}
\item At short distances the attractive/repulsive behavior is the same as in the tensionless case. Hence, identical rods will bind side-to-side and rods of opposite curvature tip-to-tip, as previously.

\item The range of the interaction is shorter in the presence of tension: the black curves decay more rapidly that the gray ones.

\item The interaction in the tip-to-tip configuration is not monotonic in the strong tension case (see Fig.~\ref{fig:th-id}b, black lines).
\end{enumerate}

It is well known that tension effects are negligible on length-scales smaller than $\xi$ while they dominate on length-scales larger than $\xi$. Accordingly, in the weak tension regime, the full interaction exhibits a crossover at $R\approx3\xi$ from a bending-dominated regime to a tension-dominated regime:
\begin{align}
H_\mathrm{int}(R)=
\begin{cases}
\tilde H_\mathrm{int}^{(0)}(R),\quad \text{for~}r_\mathrm{rod}<R\ll\xi,
\\
H_\mathrm{int}^{(1)}(R),\quad \text{for~}R\gg\xi\text{~if~}\cos[2(\theta_1-\theta_2)]\ne0,
\\
H_\mathrm{int}^{(2)}(R),\quad \text{for~}R\gg\xi\text{~if~} \cos[2(\theta_1-\theta_2)]=0,
\end{cases}
\end{align}
where
\begin{align}
H_\mathrm{int}^{(1)}(R)&=-\frac{64\pi r_\mathrm{rod}^4\xi^2}{3R^4}\kappa\, C_{\rm r1}C_{\rm r2}
\cos[2(\theta_1-\theta_2)],
\\
H_\mathrm{int}^{(2)}(R)&=
\frac{2\sqrt{2}\,\pi^{3/2}r_\mathrm{rod}^4}{9\xi^{3/2}}
\frac{e^{-R/\xi}}{\sqrt{R}}\kappa\, C_{\rm r1}C_{\rm r2}
\nonumber\\
&~~~\times
\left[
2+2\cos(2\theta_1)+2\cos(2\theta_2)+\cos(2\theta_1+2\theta_2)
\right].
\end{align}
In Fig.~\ref{fig:th-id}d, e and f, one can see the corresponding crossover from a $\sim\!1/R^2$ power-law to a $\sim\!1/R^4$ power-law.
Note that the asymptotic interaction $H_\mathrm{int}^{(1)}(R)$ depends only on the relative orientation of the inclusions, not on the direction of the separation vector. 
For particular orientations such that $\cos[2(\theta_1-\theta_2)]=0$ (graph not shown), the $\sim\!1/R^4$ power-law disappears and is replaced the exponential decay $H_\mathrm{int}^{(2)}(R)$.

\subsection{Interactions between rod assemblies}\label{sec:Tbars}

Since identical rods attract strongly and align side-to-side, while opposite rods align preferentially  tip-to-tip, we expect the following scenario. Identical rods should aggregate into straight rod assemblies, and parallel rod assemblies should either repel one another, if they are made of rods of same curvature, or attract one another side-to-side if they are made with rods of opposite curvatures. Alternate and periodic stripe structures are therefore expected to develop.

We detail in Appendix~\ref{sec:apthe2} the calculation of the interaction $F_\mathrm{int}(R)$ between two coarse-grained parallel rod assemblies of length $L$ that are separated by a center-to-center distance $R$. The result is
\begin{align}\label{intbars}
F_\mathrm{int}(R)=\frac{L}{2}\sqrt{\kappa\gamma}\,r_\mathrm{rod}^2C_\mathrm{r1}C_\mathrm{r2}\,e^{-(R-r_\mathrm{rod})/\xi}+F_\mathrm{tip},
\end{align}
where $F_\mathrm{tip}$ is the contribution coming from the extremities of the rods, which is subdominant and which we do not evaluate. 

This interaction confirms that parallel rod assemblies of like curvatures repel while rod assemblies of opposite curvatures attract. The first term, proportional to the length of the rods, vanishes in the absence of tension, i.e., for $\gamma=0$. This can easily be understood as flat membrane patches can fit in between the rod assemblies at no cost in the absence of tension. However, some interaction will originate from the extremities of the rod assemblies ($F_\mathrm{tip}$), similar to that of isolated rods. In the presence of tension, the energy per unit length will dominate and produce the attraction of rod assemblies of opposite curvature. We see then two properties: (1) the interaction is short-ranged, decaying over $\xi$, (2) the larger the tension the stronger the interaction.

\section{Simulation Model and Method}\label{sec:method}

Since the details of the meshless membrane model and protein rods are described 
in Ref.~\citenum{shib11} and Refs.~\citenum{nogu14,nogu16a}, respectively, we briefly describe the model here.
A fluid membrane is represented by a self-assembled one-layer sheet of $N$ particles.
The position and orientational vectors of the $i$-th particle are ${\bm{r}}_{i}$ and ${\bm{t}}_i$, respectively.
The membrane particles interact with each other via a potential $U=U_{\rm {rep}}+U_{\rm {att}}+U_{\rm {bend}}+U_{\rm {tilt}}$.
The potential $U_{\rm {rep}}$ is an excluded volume interaction with a diameter $\sigma$ for all pairs of particles.
Solvent is implicitly accounted for by an effective attractive potential $U_{\rm {att}}$. 

 The bending and tilt potentials
are given by 
 $U_{\rm {bend}}/k_{\rm B}T=(k_{\rm {bend}}/2) \sum_{i<j} ({\bm{t}}_{i} - {\bm{t}}_{j} - C_{\rm {bd}} \hat{\bm{r}}_{i,j} )^2 w_{\rm {cv}}(r_{i,j})$
and $U_{\rm {tilt}}/k_{\rm B}T=(k_{\rm{tilt}}/2) \sum_{i<j} [ ( {\bm{t}}_{i}\cdot \hat{\bm{r}}_{i,j})^2
 + ({\bm{t}}_{j}\cdot \hat{\bm{r}}_{i,j})^2  ] w_{\rm {cv}}(r_{i,j})$, respectively,
where ${\bm{r}}_{i,j}={\bm{r}}_{i}-{\bm{r}}_j$, $r_{i,j}=|{\bm{r}}_{i,j}|$,
 $\hat{\bm{r}}_{i,j}={\bm{r}}_{i,j}/r_{i,j}$, $w_{\rm {cv}}(r_{i,j})$ is a weight function,
and  $k_{\rm B}T$ denotes the thermal energy.
The spontaneous curvature $C_0$ of the membrane is 
given by $C_0\sigma= C_{\rm {bd}}/2$. \cite{shib11}
In this study, $C_0=0$ and $k_{\rm {bend}}=k_{\rm{tilt}}=10$ except for the membrane particles belonging to the protein rods. 

A protein rod is modeled as a linear chain of $N_{\rm {sg}}$ membrane particles.
We use $N_{\rm {sg}}=10$, which corresponds to the typical aspect ratio of the BAR domains.
The BAR domain width is approximately $2$ nm, and the length ranges from $13$ to $27$ nm. \cite{masu10}
Two types of  protein rods, called rods 1 and 2, are used.
As detailed below, rods 1 and 2 have positive and negative spontaneous curvatures $C_{\rm {r1}}$ and $C_{\rm {r2}}$ along the rod axis, respectively,
and both rods have no spontaneous curvature perpendicular to the rod axis.
Hereafter, we call the membrane particles forming a protein rod \textit{protein particles}.
The protein particles in each protein rod
 are further connected by a bond potential $U_{\rm {rbond}}/k_{\rm B}T = (k_{\rm {rbond}}/2\sigma^2)(r_{i+1,i}-\ell_{\rm sg})^2$
where $k_{\rm {rbond}}=40$, $k_{\rm {rbend}}=4000$, and $\ell_{\rm {sg}}=1.15\sigma$ are used.
The bending potential is given by $U_{\rm {rbend}}/k_{\rm B}T = (k_{\rm {rbend}}/2)(\hat{\bm{r}}_{i+1,i}\cdot\hat{\bm{r}}_{i,i-1}- C_{\rm b})^2$,
where $C_{\rm b}=1- (C_{\rm {r1}}\ell_{\rm sg})^2/2$ and  $C_{\rm b}=1- (C_{\rm {r2}}\ell_{\rm sg})^2/2$ for the rod 1 and 2, respectively.
For bonded pairs of protein particles, we use
four times larger values of $k_{\rm {bend}}$ and $k_{\rm {tilt}}$, together with the corresponding spontaneous curvature, to prevent the rod from bending tangentially in the membrane.

We employ the parameter sets used in Ref.~\citenum{nogu14}.
The membrane has mechanical properties that are typical for lipid membranes:
bending rigidity $\kappa/k_{\rm B}T=15 \pm 1$,
area of the tensionless membrane per particle $a_0/\sigma^2=1.2778\pm 0.0002$,
area compression modulus $K_A\sigma^2/k_{\rm B}T=83.1 \pm 0.4$,
and
edge line tension $\Gamma\sigma/k_{\rm B}T= 5.73 \pm 0.04$.
This edge tension $\Gamma$ is sufficiently large to prevent membrane rupture in this study~\cite{nogu16a}.
Molecular dynamics with a Langevin thermostat is employed~\cite{shib11,nogu11}.
In the following, the results are displayed with the rod length $r_{\rm {rod}}=10\sigma$ for the length unit,
 $k_{\rm B}T$ for the energy unit, and $\tau= r_{\rm {rod}}^2/D$ for the time unit,
where $D$ is the diffusion coefficient of the membrane particles in the tensionless membrane~\cite{nogu16a}.

For flat membrane simulations, the $N\gamma L_zT$ ensemble with periodic boundary conditions is used.
The projected area $A_{xy}=L_xL_y$ is fluctuated for a constant surface tension $\gamma$ while maintaining the aspect ratio $L_x=L_y$~\cite{fell95,nogu12}.
To investigate the pair correlation of the rods, two rods are set on a flat membrane with $N=6400$ and
the distance $r_{\rm {gg}}$ of the centers of the mass of the rods are constrained by a harmonic potential $k_{\rm gg}(r_{gg} -r_{\rm {rod}})^2/2$ where $k_{\rm gg}=10k_{\rm B}T/\sigma^2$. The normalized rod end-to-end vector is used to determine the rod orientation.

For self-assembly, 
the rod-1 curvature is fixed as $C_{\rm {r1}}r_{\rm {rod}}=4$, which
can induce membrane tubulation 
with a circumference of $\simeq 2r_{\rm {rod}}$ for the present bending elastic constants.
Flat membranes with $N=25600$ are investigated with various values of $C_{\rm {r2}}=-c_{\rm r}C_{\rm {r1}}$.
The density of the rods 1 and 2 are set to $\phi_{\rm {r1}}=N_{\rm {r1}}N_{\rm {sg}}/N=0.1$ and $\phi_{\rm {r1}}=N_{\rm {r2}}N_{\rm {sg}}/N=0.2$, 
where $N_{\rm {r1}}$ and  $N_{\rm {r2}}$ are the numbers of the rods 1 and 2, respectively.

For the rod assembly on a vesicle, the $NVT$ ensemble is used at  $N=9600$ and $\phi_{\rm {r1}}=\phi_{\rm {r2}}=0.25$.
The radius of the vesicle is $R_{\rm {ves}}=  3.07 r_{\rm {rod}}$ in the absence of the rods.
For the annealing simulations, the rod curvatures are changed from $C_{\rm {r1}}=C_{\rm {r2}}=0$ to $C_{\rm {r1}}r_{\rm {rod}}=4$ while maintaining the ratio $C_{\rm {r2}}=-c_{\rm r}C_{\rm {r1}}$ with annealing rate $C_{\rm {r1}}/dt=0.02/r_{\rm {rod}}\tau$ 
and subsequently the vesicle is equilibrated for $1500\sim 2500\tau$.
For each simulation condition, the reproducibility is confirmed at least from four different initial conformations.

\section{Comparison of theory and simulation on interaction of two rods}\label{sec:2rods}

Before investigating the rods self-assembly, we compare the simulation results with the theoretical predictions in the tensionless case. Due to the smallness of the interactions, it is difficult to determine numerically the potential of mean force between two rods. Instead, we compute the angular distributions and angular correlations for an isolated pair of rods separated by the short distance $R=r_{\rm rod}$. We thus determine the statistics of the following two quantities $S_1=\cos^2(\theta_1)$ and $S_2=\sin^2(\theta_1)\sin^2(\theta_2)$, as they turn out to best capture the angular behaviors (see Fig.~\ref{fig:geom} for the definition of the angles).

\begin{figure}
\centerline{\includegraphics{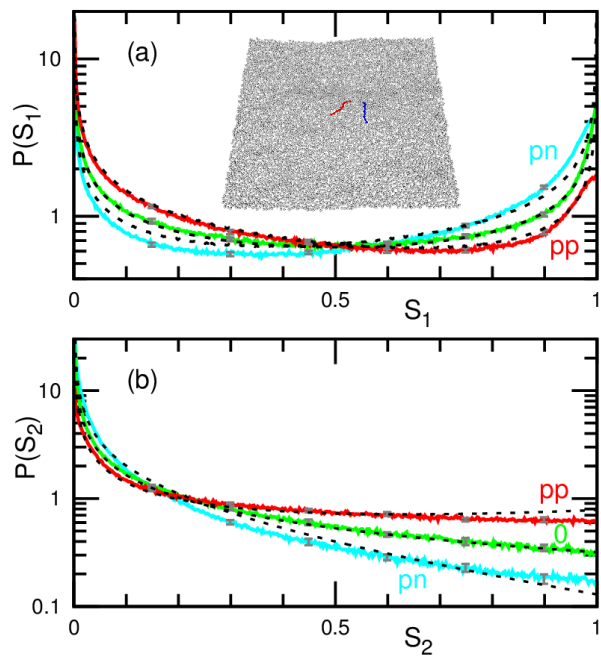}}
\caption{
Probability distribution of the angular parameters $S_1$ (a) and $S_2$ (b) for two rods separated by $R=r_{\rm {rod}}$  
at zero surface tension ($\gamma=0$).
The colored solid lines represent the simulation data 
for identical positive rod curvature $C_{\rm {r1}}=C_{\rm {r2}}=4/r_{\rm {rod}}$ (red, labeled by pp),
opposite curvatures $C_{\rm {r1}}= -C_{\rm {r2}}=4/r_{\rm {rod}}$ (cyan, labeled by pn), and zero curvature $C_{\rm {r1}}= C_{\rm {r2}}=0$ (green, labeled by 0).
The dashed lines are deduced from $\tilde H_\text{int}$ in eqn~(\ref{eq:Hint}), taking a prefactor of 
$3 k_{\rm B}T$, $-3 k_{\rm B}T$, and $0$ for pp, pn, and 0, respectively.
A typical snapshot is shown in the inset of (a). 
The red and blue particles represent the rod segments and the gray particles represent membrane particles.
}
\label{fig:angpair}
\end{figure}

The theory predicts $P(\theta_1,\theta_2)=\exp(-\tilde H_\text{int}/k_{\rm B}T)/Z$, with $\tilde H_\text{int}$ given by eqn~(\ref{eq:Hint}) where  $Z$ is the normalization factor for $\int_0^1 P(S_a)dS_a=1$ ($a=1$ or $2$).
For rods with zero curvatures, $\tilde H_\text{int}=0$, thus $P(\theta_1)=1/(2\pi)$ and $P(S_1)=1/[\pi\sqrt{S_1(1-S_1)}]$.
The simulations confirm  this dependence, as shown by the green and superimposed dashed lines in Figs.~\ref{fig:angpair}a and~b. This  shows that if there is a Casimir-like fluctuation--induced interaction (neglected in our calculations), arising from the contrast of rigidity between the rods and the surrounding membrane, it must be very weak.

We consider now curved rods. When two rods have equal curvatures, the theory implies that the orientations $\bm{u}_1$ and $\bm{u}_2$ are preferentially perpendicular to $\bm{r}_{12}$ and therefore smaller values of $S_1$ become more probable. Conversely, for rods of opposite curvatures, $\bm{u}_1$ and $\bm{u}_2$ are preferentially parallel to $\bm{r}_{12}$ and thus larger values of $S_1$ become more probable. These trends are confirmed by the simulations, as evidenced by the red and cyan lines in Fig.~\ref{fig:angpair}a. For the same reasons, large values of $S_2$ have larger probability for rods of equal curvatures and smaller probability for rods of opposite curvatures, in agreement with Fig.~\ref{fig:angpair}b. With the parameters $\kappa$, $r_\mathrm{rod}$ and $R$ given previously, we obtain a good fit of the numerical results by the theory provided we renormalize the curvatures by a factor $\simeq\!1/20$ (see the agreement between the dashed lines and the colored solid lines in Figs.~\ref{fig:angpair}a and~b). This apparently large factor is reasonable given the amplitude of the rod shape fluctuations in Fig.~\ref{fig:angpair}a, and it is probably due to the finite values of $k_\mathrm{rbond}$ and $k_\mathrm{rbend}$ and to the large number of particles in a rod ($N_\mathrm{sg}=10$). We conclude that our model captures quantitatively the angular dependence and the amplitude of the rods interaction provided renormalized curvatures are used.

\section{rod assembly on flat membrane}\label{sec:flat}

\begin{figure}
\centerline{\includegraphics{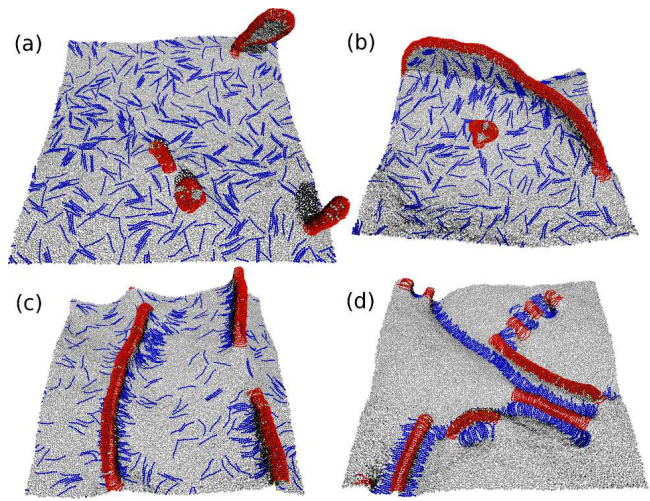}}
\caption{
Snapshots of assembled protein rods on the membrane at the surface tension $\gamma=0$.
(a) the ratio of the rod curvatures, $c_{\rm r}=-C_{\rm {r2}}/C_{\rm {r1}}=0$.
(b) $c_{\rm r}=0.25$.
(c) $c_{\rm r}=0.5$.
(d) $c_{\rm r}=0.75$.
The red and blue particles represent the segments of the rod 1 and 2, respectively.
The gray particles represent membrane particles.
}
\label{fig:snap_cs}
\end{figure}

\begin{figure}
\centerline{\includegraphics{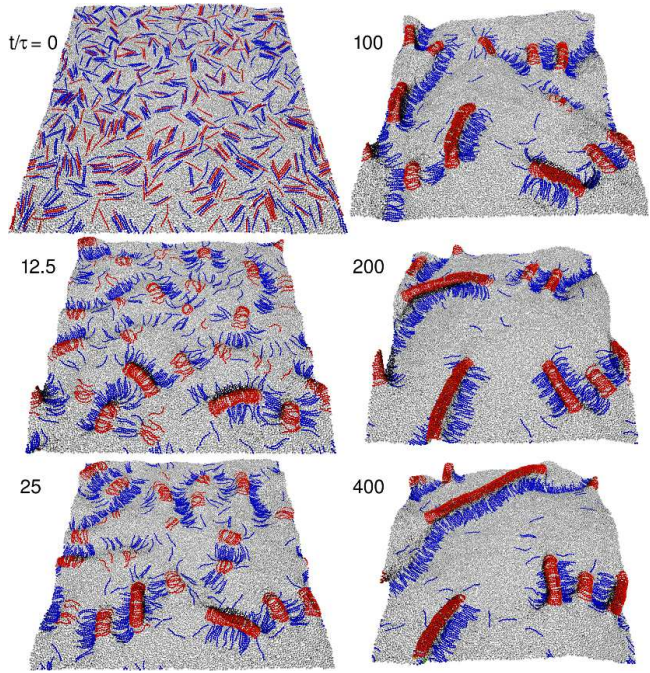}}
\caption{
Sequential snapshots of self-assembly of protein rods at the rod-curvature ratio $c_{\rm r}=0.625$ 
and the surface tension $\gamma=0$.
}
\label{fig:snap_cn5}
\end{figure}

First, we describe the rod assemblies on the tensionless flat membrane ($\gamma=0$).
Figure~\ref{fig:snap_cs} shows typical snapshots.
For zero rod-curvature of the rods 2 ($c_{\rm r}=0$), the rods 1 form disk-shaped tubules like in the absence of rods 2.
The rods 2 are distributed on the flat region of the membrane.
Small tubules have a semicircular-disk shape.
In contract, large tubules have a disk shape (like a mussel shell) connected to the flat membrane by a narrow cylindrical neck 
(see the upper tubule in Fig.~\ref{fig:snap_cs}a).
We believe that these disk-like structures could be the microscopic precursors of the tubules induced by BAR proteins in experiments. Like in other endocytotic processes (e.g., clathrin--mediated endocytosis) the rods shaping these tubular vesicles are at the same time their cargos.
This assembly of the rods 1 agrees with the theoretical prediction for the point-like inclusions in Sec.~\ref{sec:Tpoint} (identical curved rods attract and align side-to-side).
Identical rods have an attractive interaction perpendicular to the rod axes.
The neighboring rods 1 contact each other 
and subsequently many rods 1 assemble and form a straight one-dimensional band curved along the short axis of the assembly (parallel to the rod axis).
Eventually, long band assemblies bend along their long axis and form a tubule, as this reduces the bending energy cost around the assembly.
The rods 2, with $C_{\rm {r2}}=0$, exhibit little interactions with the rods 1.
This also agrees with the prediction of eqn~(\ref{eq:Hint}).

As $c_{\rm r}$ increases, the necks of the large tubules remain open.
For $c_{\rm r}=0.25$, 
the rod-1 assemblies exhibit wide walls or hill shapes (see Fig.~\ref{fig:snap_cs}b).
The rods 2 are concentrated around the foot of these walls and  stabilize their negative curvature.
With further increasing $c_{\rm r}$,
the rods 2 more clearly assemble at both sides of the rod-1 assemblies 
and form long straight bumps [see Fig.~\ref{fig:snap_cs}c and the corresponding movie provided in ESI (Movie 1)].
These structures recall the shape of a centipede or myriapod:
the rod-1 straight assembly looks like the body of a centipede and the rods 2 on both sides are like many legs.
This attraction of oppositely curved rods along their axis also agrees with the theoretical prediction 
for the point-like inclusions in Sec.~\ref{sec:Tpoint} (oppositely curved rods attract and align tip-to-tip).

The straight assemblies are also stable at larger $c_{\rm r}$.
However, at $c_{\rm r}\gtrsim 0.58$, 
straight assemblies of opposite curvature have attractive interaction in the lateral direction (parallel to the axis of each rod)
and form a periodic bump structure (see Fig.~\ref{fig:snap_cs}d).
Each rod-1 (-2) straight assembly is connected to two rod-2 (-1) straight assemblies on both lateral sides.
We call this assembly of the rods 1 and 2 a \textit{stripe assembly},
since it forms periodic bands.
This stripe assembly is metastable but has a longer life time than our simulation periods 
at $c_{\rm r}=0.75$.
Even when the stripe assembly is set as an initial conformation,
the rod-1 straight assemblies separate to form isolated bumps at $c_{\rm r}\lesssim 0.5$.

\begin{figure}
\centerline{\includegraphics{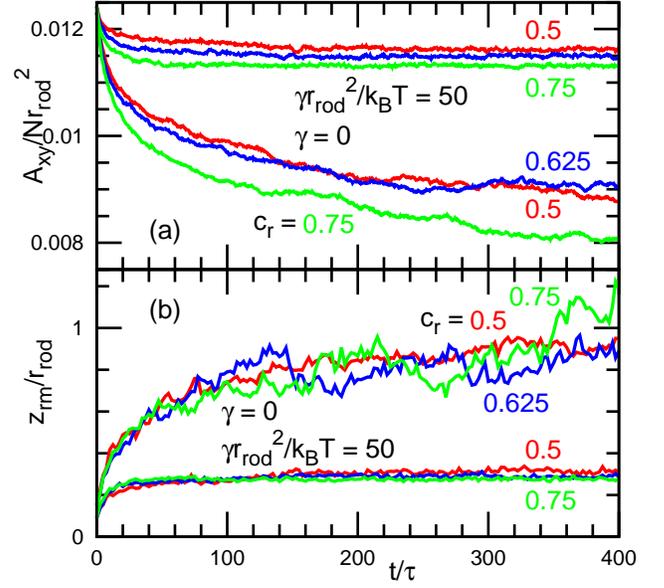}}
\caption{
Time evolution of (a) the projected membrane area $A_{xy}$ and (b) mean vertical rod span $z_{\rm rm}$ 
for the rod-curvature ratio $c_{\rm r}=0.5$, $0.625$, and $0.75$
 at the surface tension $\gamma r_{\rm {rod}}^2/k_{\rm B}T=0$ and $50$.
At $c_{\rm r}=0.625$ and $\gamma=0$, 
the same data are used as in Fig.~\ref{fig:snap_cn5}.
}
\label{fig:ar}
\end{figure}

\begin{figure}
\centerline{\includegraphics{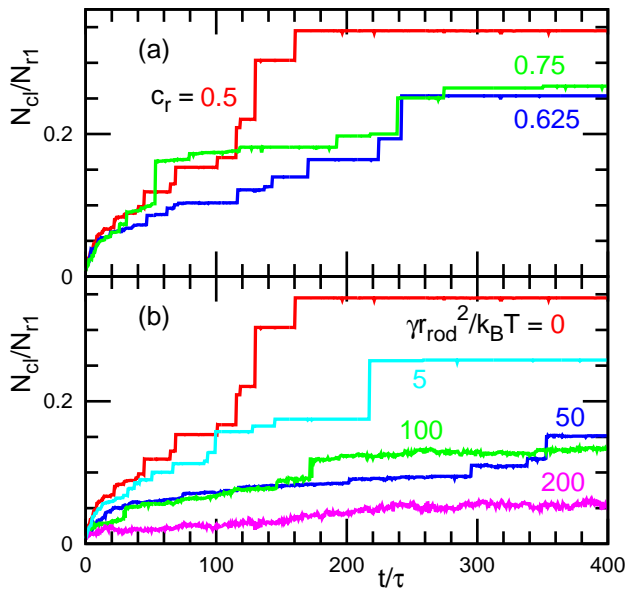}}
\caption{
Time evolution of mean cluster size of the rods 1.
(a) the rod-curvature ratio $c_{\rm r}=0.5$, $0.625$, and $0.75$ at the surface tension $\gamma=0$.
(b) $\gamma r_{\rm {rod}}^2/k_{\rm B}T=0$, $5$, $50$, $100$, and $200$ at $c_{\rm r}=0.5$.
The same data are used as in Fig.~\ref{fig:ar} for the corresponding parameter sets.
}
\label{fig:cl}
\end{figure}

\begin{figure}
\centerline{\includegraphics{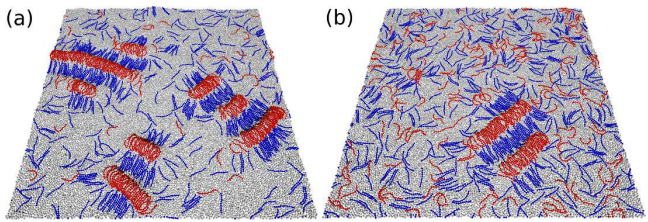}}
\caption{
Snapshots of the assembled protein rods on the  membrane at the rod-curvature ratio $c_{\rm r}=0.5$ 
for (a) the surface tension $\gamma r_{\rm {rod}}^2/k_{\rm B}T=100$ and (b) $\gamma r_{\rm {rod}}^2/k_{\rm B}T=200$.
}
\label{fig:snap_f}
\end{figure}

\begin{figure}
\centerline{\includegraphics{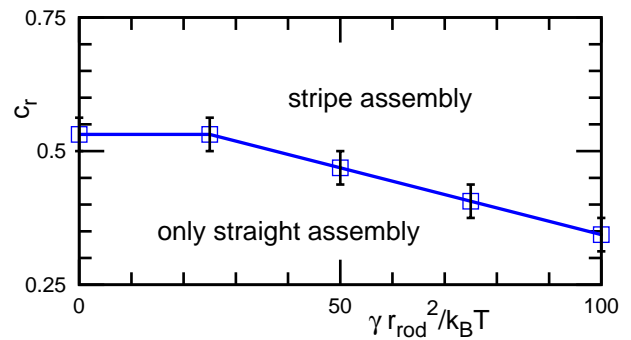}}
\caption{
Phase diagram of the rod assembly for the surface tension $\gamma$ and the ratio of the rod curvatures $c_{\rm r}$.
}
\label{fig:pd}
\end{figure}

Figure~\ref{fig:snap_cn5} shows the assembly dynamics  
at $c_{\rm r}=0.625$ [The corresponding movie is provided in ESI (Movie 2)].
First, short straight assemblies are formed. 
As the tips of the bumps approach closely, they fuse into one large straight assembly.
In contrast, as they approach laterally, they form stripe assembly of short straight assemblies.
As a long bump approaches, the stripe assemblies are disassembled and reassemble into longer assemblies (see the upper region of the three right snapshots in Fig.~\ref{fig:snap_cn5}).
Note that some rods of type 2 also contact the rod-1 assemblies at the tips of the assemblies 
(see snapshots at $t/\tau=12.5$ and $25$ in Fig.~\ref{fig:snap_cn5}),
so that an attractive interaction exists between the rods of types 1 and 2 even when the rod-1 axis is perpendicular to the separation vector.
This attraction is in good agreement with the prediction of eqn~(\ref{eq:Hint}) with $\theta_1=0$ and $\theta_2=\pi/2$.

As the rods assemble, the projected membrane area $A_{xy}$ decreases 
while the mean vertical rod span $z_{\rm rm}$ and the mean cluster size $N_{\rm {cl}}$ of rods 1 increase 
(see Figs.~\ref{fig:ar} and \ref{fig:cl}).
At $c_{\rm r}\geq 0.625$, the rod-1 cluster formation becomes slower 
than at $c_{\rm r}= 0.5$, since the stripe assembly suppresses the fusion between the rod-1 assemblies.
The vertical span is calculated from 
the height variance of all rod segments as 
$z_{\rm rm}^2=\sum_{i\in {\rm rods}} (z_i-z_{\rm G})^2/(N_{\rm {r1}}+N_{\rm {r2}})N_{\rm {sg}}$
where $z_{\rm G}=\sum_{i\in {\rm rods}} z_i/(N_{\rm {r1}}+N_{\rm {r2}})N_{\rm {sg}}$.
A rod is considered to belong to a cluster
when the distance between the centers of mass of the rod 
and one of the rods in the cluster is less than $r_{\rm {rod}}/2$. 
For the rods 1, the mean cluster size is given by
$N_{\rm {cl}}=  (\sum_{i_{\rm {cl}}=1}^{N_{\rm {r1}}} i_{\rm {cl}}^2 n^{{\rm {cl}}}_i)/N_{\rm {r1}}$
with $N_{\rm {r1}}=\sum_{i_{\rm {cl}}=1}^{N_{\rm {r1}}} i_{\rm {cl}} n^{{\rm {cl}}}_i$
where $n^{{\rm {cl}}}_i$  is the number of clusters of size $i_{\rm {cl}}$.

At small positive surface tensions ($\xi \gg r_{\rm {rod}}$, {\it i.e.}, $\gamma \ll \gamma_{\rm c}=\kappa/r_{\rm {rod}}^2$),
the bending energy contribution should be dominant in the interactions between the rods as described in Sec.~\ref{sec:theory}.
Our simulation results support this theoretical prediction.
The rod assembly is only slightly modified at $\gamma r_{\rm {rod}}^2/k_{\rm B}T \lesssim 5$ ($\xi \gtrsim 1.7r_{\rm {rod}}$).
In contrast, a large positive surface tension ($\gamma\gtrsim \gamma_{\rm c}$) suppresses both
the decrease of the projected area $A_{xy}$
and the protrusion of the rod assemblies (see Fig.~\ref{fig:ar}).
Since the curvature-energy gain by the rod assembling is reduced when increasing $\gamma$ (see the theory in Sec.~\ref{sec:theory_tension} and Fig.~\ref{fig:th-id}),
the rod cluster size decreases and more rods remain in an isolated state
(see Figs.~\ref{fig:cl}b and \ref{fig:snap_f}).
Interestingly, the stripe assembly is stabilized for smaller $c_{\rm r}$
at larger values of $\gamma$, as shown in Fig.~\ref{fig:pd}.
For $c_{\rm r}=0.5$, the stripe assembly  exists for $\gamma r_{\rm {rod}}^2/k_{\rm B}T \gtrsim 50$
while it does not for $\gamma r_{\rm {rod}}^2/k_{\rm B}T \lesssim 50$ (compare Figs.~\ref{fig:snap_cs}c and \ref{fig:snap_f}). This agrees well with the theoretical results of Sec.~\ref{sec:Tbars} showing that the attraction between long straight assemblies of opposite curvature scales as $\sim\!\gamma^{1/2}$ and thus increases with  membrane tension.

\begin{figure}
\includegraphics{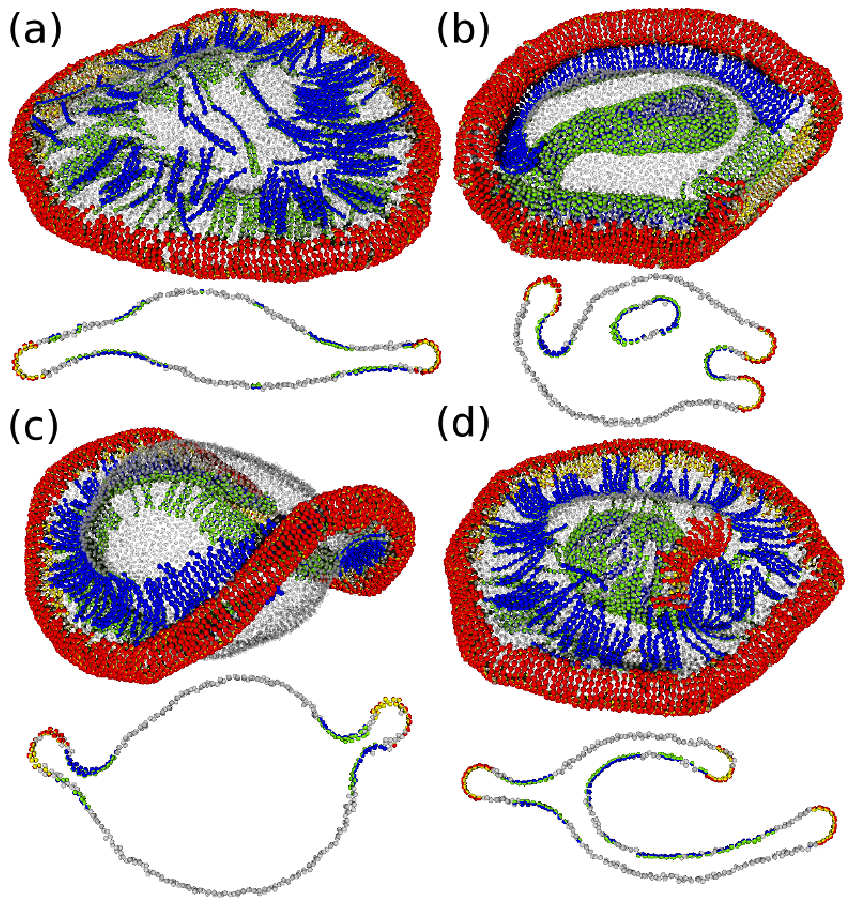}
\includegraphics{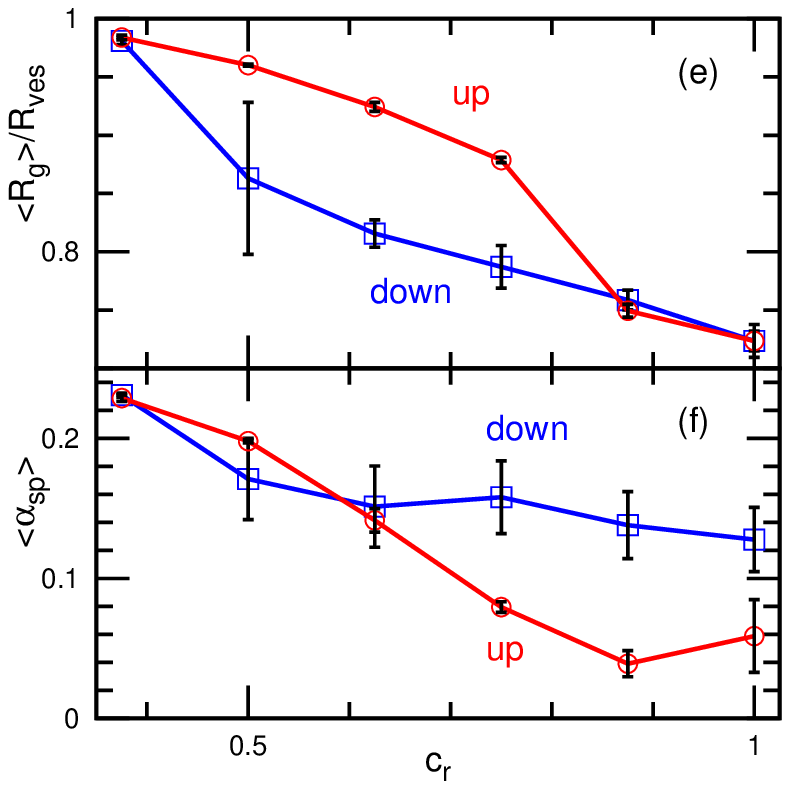}
\caption{
Vesicle deformation by oppositely curved  rods at various rod-curvature ratios $c_{\rm r}$. (a)--(d) Snapshots. The membrane particles are represented by a transparent gray sphere.
The rod 1 (resp.\ rod 2) segments are represented by a sphere half
red and half yellow (resp.\ half blue and half green).
The orientation vector ${\bm{t}}_i$ lies along the direction from the 
yellow (green) to red (blue) hemispheres.
(a), (b) Vesicles formed by annealing to $C_{\rm {r1}}r_{\rm {rod}}=4$
with (a) $c_{\rm r}=0.375$ and (b) $c_{\rm r}=1$. 
(c) Vesicles formed by an increase in $c_{\rm r}$ from $0.375$ to $0.75$ at $C_{\rm {r1}}r_{\rm {rod}}=4$.
(d) Vesicles formed by a decrease in $c_{\rm r}$ from $1$ to $0.5$ at $C_{\rm {r1}}r_{\rm {rod}}=4$.
Top and bottom panels show the snapshot of all particles in bird's-eye view and a sliced snapshot in front view, respectively. 
(e),(f) Dependence of (e) the mean radius of gyration $\langle R_g\rangle$ and 
(f) mean asphericity $\langle\alpha_{\rm {sp}}\rangle$
of vesicles on $c_{\rm r}$.
The data labeled up and down are obtained by an increase and decrease in $c_{\rm r}$ 
at $C_{\rm {r1}}r_{\rm {rod}}=4$.
}
\label{fig:snap_ves}
\end{figure}

\section{Vesicle deformation by rod assembly}\label{sec:ves}

The rods also assemble on vesicles like on flat membranes.
However, the assembly structures are largely modified by the original vesicle curvature
and its closed geometry.
When a single type of rod exists on the vesicle,
the rods induce oblate or polyhedral shapes of the vesicle and they assemble 
at the edges of the polyhedron~\cite{nogu14,nogu15b}.
For negative rod curvatures, invaginations into the inside of the vesicle can occur~\cite{nogu15b,nogu16}.
Here, we set the rod-1 density, $\phi_{\rm {r1}}=0.25$, to make the vesicle form
a disk shape in the absence of the rods 2.
We add the rods 2 with the same density and investigate how the vesicle shape is changed upon varying $c_{\rm r}$.
Note that the vesicles maintain the spherical topology without membrane rupture.

Figure~\ref{fig:snap_ves}a and b show the snapshots obtained 
by annealing from $C_{\rm {r1}}=C_{\rm {r2}}=0$ to $C_{\rm {r1}}r_{\rm {rod}}=4$ at fixed ratio $c_{\rm r}$.
For $c_{\rm r}=0.375$,
the vesicles form a disk shape with a concave circular region between the center and the rim;
The rods 2 are aligned along radial directions of the disk in this concave region.
For $c_{\rm r}=1$,
the rods 2 form invaginations into the inside of the vesicle
and the two rims of the rod-1 assembly are partially connected by the rod-2 assembly like in the flat membrane.
For intermediate values of $c_{\rm r}$, 
the obtained vesicle shapes depend on the initial conformations and thermal noise during the annealing:
Some vesicles have an invagination or partial doubled rims.

To investigate the stability of the disk-shaped and invaginated vesicles,
we simulated vesicles with different $c_{\rm r}$ starting from the annealed conformation 
at $c_{\rm r}=0.375$ and $1$ as initial states.
As $c_{\rm r}$ increases for disk-shaped vesicles,
the center region of the vesicle becomes more spherical.
This shape change reduces the radius of the disk,
resulting in a buckling or winding of the disk rim, at $c_{\rm r}=0.625$ and $0.75$
(see Fig.~\ref{fig:snap_ves}c).
At $c_{\rm r}=0.875$ and $1$, 
the rods 2 form a straight dimple on the side of the rod-1 assembly (as seen in the left region of the sliced snapshot in Fig.~\ref{fig:snap_ves}b) 
as well as an invagination.

Starting from an invaginated vesicle, 
the invagination swells to an ellipsoidal shape
and the rest of the rods 1 form a circular rim, at $c_{\rm r}=0.5$ (see Fig.~\ref{fig:snap_ves}d).
At $c_{\rm r}=0.375$, the invagination is removed
and a disk-shaped vesicle is obtained [see the movie is provided in ESI (Movie 3)].

These shape changes can be characterized by the radius of gyration $R_{\rm g}$ and 
the shape parameter, asphericity $\alpha_{\rm {sp}}$ (see Fig.~\ref{fig:snap_ves}e and f).
The asphericity is the degree of deviation from a spherical 
shape and is expressed as~\cite{rudn86}
\begin{equation}
\alpha_{\rm {sp}} = \frac{({\lambda_1}-{\lambda_2})^2 + 
  ({\lambda_2}-{\lambda_3})^2+({\lambda_3}-{\lambda_1})^2}{2 R_{\rm g}^4},
\end{equation}
where $R_{\rm g}^2={\lambda_1} + {\lambda_2} + {\lambda_3}$ and
${\lambda_1} \leq {\lambda_2} \leq {\lambda_3}$ are the 
eigenvalues of the gyration tensor of the vesicle:
 $a_{\alpha\beta}= \sum_j (\alpha_{j}-\alpha_{\rm G})(\beta_{j}-\beta_{\rm G})/N$,
where $\alpha, \beta=x,y,z$ and $\alpha_{\rm G}$ are the coordinates of the center of mass.
For a perfect sphere, $\alpha_{\rm {sp}} = 0$, for a thin rod $\alpha_{\rm {sp}}=1$, 
 and for a thin disk, $\alpha_{\rm {sp}}=0.25$.
Starting from a disk-shaped vesicle (labeled `up' in Fig.~\ref{fig:snap_ves}e and f),
the vesicles have more spherical and less compact shapes than invaginated vesicles 
so that they take larger $R_{\rm g}$ and smaller $\alpha_{\rm {sp}}$.
Thus, the vesicle shapes can be chosen from many metastable conformations
 by using the hysteresis of $c_{\rm r}$ variations. Our simulations demonstrate 
a  control method of the vesicle shapes by using hysteresis.

\section{Summary and discussion}\label{sec:sum}

We have studied the assembly structures of binary banana-shaped protein rods.
Compared to the situation where only one  type of rod is present,
the coexistence of two types of rods, with opposite curvatures, induces a greater variety of membrane shapes. Here, using meshless membrane simulations, we demonstrated straight bumps and stripe structures in flat membranes,
and buckled rims and ellipsoidal invaginations in vesicles.
More complicated shapes such as vesicles with periodic bumps, like cartridge pleats,
can be formed for larger vesicles.
These structure formations can be examined in experiments with two types of BAR proteins absorbed on the same side of the membrane
or with a single type of protein absorbed on both side of the membrane.
The periodic length of the stripe structure is the sum of the longitudinal lengths of two BAR domains ($20$ to $50$ nm). It should therefore be observable by electron microscopy.
Our simulation suggests that imposing a positive surface tension helps to obtain the stripe structure.

For rods of like curvatures, our theory
predicts a side-to-side attractive interaction
and a tip-to-tip repulsion. For rods having curvatures of opposite sign our theory predicts the reverse, i.e., a tip-to-tip attraction and a side-to-side repulsion.
Our coarse-grained simulations revealed that 
identical rods assemble in the side-to-side configuration and thus build straight rod assemblies; these assemblies attract the rods with opposite curvatures in the tip-to-tip configuration thereby inducing a side-to-side alignment of the formed assemblies.
The formation of those alternate assemblies, or stripe structures, therefore agrees well with our theoretical predictions.

With increasing membrane tension, we theoretically found that the attractive interaction between opposite rods in the tip-to-tip configuration (which corresponds to the orientations involved in the stripe structures) is increased at short distances and decreased at large distances (see Fig.~\ref{fig:th-id}b). Likewise, our larger-scale analysis of Sec.~\ref{sec:Tbars} predicts a short-range attraction between opposite straight rod assemblies that increases with membrane tension. These theoretical results explain well the observed increased stability of the stripe structures when the tension is increased and the shorter length of the rod assemblies in the thermalized system.

These theoretical predictions are supported by our simulation results although
the theory considers rigid rods with a fixed shape while the rods in the simulation are flexible. Our model captures quantitatively the angular dependence and the amplitude of the rods interaction provided renormalized curvatures are used.
The real proteins are not completely rigid but their stiffness is not measured.
The rods flexibility can be considered to result in an effectively smaller rod curvature, not altering the sign of the interactions in the present assemblies.
The protein stiffness, together with the finite size of the rod inclusions, should however be taken into account for quantitative predictions of the assembly conditions.

Here, we consider the case where the protein rods have zero spontaneous (side) curvature perpendicular to the rod axis.
The protein--membrane interactions including amphipathic-helix insertions can yield non-zero side curvature.
The finite rod and side curvatures can induce egg-carton~\cite{domm99,domm02} and network structures~\cite{nogu16}.
If the membrane bending in perpendicular direction is much stronger than that along the rod axis,
the effective anisotropic axis of the curvature is perpendicular to the rod axis.
In such a case, the side-to-side attraction of the anisotropic inclusions corresponds 
the tip-to-tip attraction for the rods axes.
The linear assembly of the N-BAR domains in the coarse-grained molecular simulation of Ref.~\citenum{simu13} 
may be understood by the effect of the strong side curvature.
For quantitative comparison of our results with experiments, the elastic parameters of each type of protein are required.
The estimation of the protein stiffness and side curvature by atomic or coarse-grained molecular simulation is important.

It is interesting also to note that we find a good agreement between theory and simulations although our calculations are set in the linearized regime for the membrane deformation, contrary to what was found isotropic particles in the strong deformation regime~\cite{reyn07}. Either anisotropic particles are more adapted to linearized calculations, or the curvatures in the regimes where the assemblies take place are not so strong: further investigations along these lines would be interesting.

The effects of the Casimir force, induced by the fluctuations of the membrane, remains to be investigated further. 
It was ignored in the present work since it is normally sub-dominant for small inclusions imposing large membrane deformations.
This Casimir force is known however to induce an effective attraction between long straight rods~\cite{Golestanian96PRE,Bitbol11EPL}.
We have investigated straight rods in our simulations, and found that the latter have too weak Casimir interactions to induce rod clusters for the present length ($N_{\rm {sg}}=10$).
However, twice longer rods ($N_{\rm {sg}}=20$) can form clusters.
Thus, the assembly of longer rods can be induced by both Casimir and rod-curvature forces.

In living cells, many types of BAR superfamily proteins cooperate
to regulate membrane shapes.
Here, we have studied only the mixture of two types of proteins.
The cooperative effect of mixing more than two proteins is an important topic for further studies.

\section*{Acknowledgments}
This work was supported by JSPS KAKENHI Grant Number JP25103010
and MEXT as ``Exploratory Challenge on Post-K computer'' (Frontiers of Basic Science: Challenging the Limits).
Numerical calculations were partly carried out on SGI Altix ICE 
XA at Supercomputer Center of ISSP, University of Tokyo. 

\begin{appendix}

\section{Interactions between point-like inclusions constraining the membrane curvature along a specific direction}\label{sec:apthe}

In the small deformation regime,
the membrane shape can be expressed in the Monge gauge by a height function $z=h(\bm{r})$, with $\bm{r}=(x,y)$ a point of the projected plane. The Helfrich elastic energy of the membrane is given, to quadratic order, by~\cite{Helfrich73}
\begin{align} \label{eq:hel}
H\simeq\int\!d^2r\left[\frac{\kappa}{2}\left(\nabla^2 h\right)^2+\frac{\gamma}{2}\left(\nabla h\right)^2\right]
=\frac{\kappa}{2}\int\!d^2r\,h\mathcal{L}h,
\end{align}
with $\mathcal{L}=\nabla^4-\xi^{-2}\nabla^2$, the operator associated with the Hamiltonian, and $\xi=\sqrt{\kappa/\gamma}$ the coherence length arising from the membrane tension $\gamma$ and the membrane bending rigidity $\kappa$.

Let us consider a point-like inclusion, placed at $\bm{r}=\bm{0}$, that imposes a membrane curvature $C_\mathrm{r1}$ along the direction given by the unit vector $\bm{u}_1$ of angle $\theta_1$ (see the definition of the angles in Sec.~\ref{sec:Tpoint}), and a second point-like inclusion, placed at $\bm{r}=R\bm{e}_x$, imposing a curvature $C_\mathrm{r2}$ along the direction given by the unit vector $\bm{u}_2$ of angle $\theta_2$. We shall discuss later on how to mathematically fix the ``size" of these inclusions.
The constraints set by the inclusions, for small membrane deformations, read
\begin{align}\label{cons1}
h_{,11}(\bm{0}) & \equiv   h_{,xx}(\bm{0})\cos^2\theta_1 + h_{,xy}(\bm{0})\sin(2\theta_1) \nonumber \\  &  +  h_{,yy}(\bm{0})\sin^2\theta_1 = C_\mathrm{r1}, 
\\\label{cons2}
h_{,22}(R\bm{e}_x) & \equiv   h_{,xx}(R\bm{e}_x)\cos^2\theta'_2 + h_{,xy}(R\bm{e}_x)\sin(2\theta'_2) \nonumber \\   &  + h_{,yy}(R\bm{e}_x)\sin^2\theta'_2 = C_\mathrm{r2},
\end{align}
where a comma indicates differentiation. Here $\theta'_2$ is the polar angle of the direction $\bm{u}_2$ with respect to the $x$-axis, i.e., $\theta'_2=\pi-\theta_2$.
Minimizing the Helfrich energy~(\ref{eq:hel})
with the above constraints yields the Euler--Lagrange equation:
\begin{align}\label{eq:Lh}
\mathcal{L}h(\bm{r})&=
\Lambda_1\Big[
\delta_{,xx}(\bm{r})\cos^2\theta_1 + \delta_{,xy}(\bm{r})\sin(2\theta_1) \\ & + \delta_{,yy}(\bm{r})\sin^2\theta_1
\Big]
+\Lambda_2\Big[
\delta_{,xx}(\bm{r}-R\bm{e}_x)\cos^2\theta'_2  \nonumber \\ & + \delta_{,xy}(\bm{r}-R\bm{e}_x)\sin(2\theta'_2) + \delta_{,yy}(\bm{r}-R\bm{e}_x)\sin^2\theta'_2
\Big]. \nonumber
\end{align}
Here, $\delta$ is the Dirac distribution, and $\Lambda_1$ and $\Lambda_2$ are Lagrange multipliers. The solution is given by
\begin{align}
h(\bm{r})&=\Lambda_1\Big[
G_{,xx}(\bm{r})\cos^2\theta_1 + G_{,xy}(\bm{r})\sin(2\theta_1)  \\ & + G_{,yy}(\bm{r})\sin^2\theta_1
\Big]
+\Lambda_2\Big[
G_{,xx}(\bm{r}-R\bm{e}_x)\cos^2\theta'_2  \nonumber \\ & + G_{,xy}(\bm{r}-R\bm{e}_x)\sin(2\theta'_2) + G_{,yy}(\bm{r}-R\bm{e}_x)\sin^2\theta'_2
\Big], \nonumber
\end{align}
where $G(\bm{r})$ is the Green function defined by
\begin{equation}\label{eq:Green}
\mathcal{L}G(\bm{r})=\delta(\bm{r})\,.
\end{equation}

\subsection{Vanishing tension case}

If $\gamma=0$, the Euler-Lagrange operator becomes $\mathcal{L}=\nabla^4$ and the Green function is given by~\cite{Goulian93EPL}
\begin{equation}\label{eq:freeGreen}
G(r)=\frac{r^2}{8\pi}\ln r.
\end{equation}
Indeed, with $\nabla_r^2=r^{-1}\partial_r r\partial_r$ (cylindrical coordinates), the most general rotationally symmetric solution of $\nabla^2G(\bm{r})=0$ is $G(r)=A_1+A_2\ln r+A_3 r^2+A_4 r^2\ln r$. The finiteness of $G$ at the origin requires $A_2=0$ and integrating eqn~(\ref{eq:Green}) around $\bm{r}=0$ yields $A_4=1/(8\pi)$. The constants $A_1$ and $A_3$ are useful if one wishes to satisfy some boundary condition for large $r$, however they do not contribute to the fourth derivatives of $G$ appearing below, so we can set $A_1=A_3=0$, which yields eqn~(\ref{eq:freeGreen}).

Satisfying the constraints (\ref{cons1}) and (\ref{cons2}) gives a linear set of equations for the Lagrange multipliers:
\begin{align}
\mathsf{M}\Lambda = \mathsf{C},
\end{align}
where $\Lambda=(\Lambda_1, \Lambda_2)^T$, $\mathsf{C}=(C_{\rm r1},C_{\rm r2})^T$ and $\mathsf{M}$ is a matrix such that
\begin{align}
M_{11}&=\label{eq:M11}
G_{,xxxx}(\bm{0})\cos^4\theta_1 +
4G_{,xxxy}(\bm{0})\cos^3\theta_1\sin\theta_1  \nonumber\\ & +
6G_{,xxyy}(\bm{0})\cos^2\theta_1\sin^2\theta_1 \\ & +
4G_{,xyyy}(\bm{0})\cos\theta_1\sin^3\theta_1   +
G_{,yyyy}(\bm{0})\sin^4\theta_1\,,  \nonumber
\\
M_{21}&=G_{,xxxx}(R\bm{e}_x)\cos^2\theta_1\cos^2\theta'_2  \nonumber\\ &+
2G_{,xxxy}(R\bm{e}_x)\cos\theta_1\cos\theta'_2\sin(\theta_1+\theta'_2) \nonumber\\ &+
\frac{1}{4}G_{,xxyy}(R\bm{e}_x)\left[2+\cos(2\theta_1-2\theta'_2)-3\cos(2\theta_1+2\theta'_2)\right]
\nonumber\\ &+
2G_{,xyyy}(R\bm{e}_x)\sin\theta_1\sin\theta'_2\sin(\theta_1+\theta'_2) \nonumber\\ &+
G_{,yyyy}(R\bm{e}_x)\sin^2\theta_1\sin^2\theta'_2\,,
\end{align}
and with $M_{12}$ obtained from $M_{21}$ by replacing $R\bm{e}_x$ by $-R\bm{e}_x$, and with $M_{22}$ obtained from $M_{11}$ by replacing $\theta_1$ by $\theta'_2$. Using eqns~(\ref{eq:hel}) and~(\ref{eq:Lh}), one obtains exactly
\begin{align}\label{Hbyparts}
H=\frac{\kappa}{2}(\Lambda_1C_\mathrm{r1}+\Lambda_2C_\mathrm{r2})=\frac{\kappa}{2}\mathsf{C}^T\mathsf{M}^{-1}\mathsf{C}.
\end{align}
We need the fourth derivatives of the Green function. From eqn~(\ref{eq:freeGreen}), we obtain
\begin{align}
&G_{,xxxx}(\pm R\bm{e}_x)=G_{,xxyy}(\pm R\bm{e}_x)=-\frac{1}{4\pi R^2}, \nonumber \\
&G_{,xxxy}(\pm R\bm{e}_x)=G_{,xyyy}(\pm R\bm{e}_x)=0,\\
&G_{,yyyy}(\pm R\bm{e}_x)=\frac{3}{4\pi R^2}, \nonumber
\end{align}
It follows that the fourth derivatives of $G(\bm{r})$, that appear in $M_{11}$ and $M_{22}$, are singular. This problem comes from the fact that we use mathematically a point-like constraint. Physically, the inclusions have a characteristic size, here $a\approx r_\mathrm{rod}$, and we can specify this size by using an upper wavevector cutoff $\Lambda\approx1/a$ for the membrane deformation in the reciprocal space. We take numerically $\Lambda=2/a$ as in Refs.~\citenum{Park96JPhysI,domm99}, since this choice was shown to match the exact calculations in the case of isotropic inclusions.
This gives
\begin{align}
G_{,xxxx}(\bm{0})=\int_0^{2/a}\!\!\frac{q\,dq}{(2\pi)^2}\int_0^{2\pi}\!\!d\theta\,\cos^4\theta
=\frac{3}{8\pi a^2}.
\end{align}
Similarly, 
\begin{align}
G_{,xxxy}(\bm{0})&= G_{,xyyy}(\bm{0}) =0,\\ \nonumber
G_{,xxyy}(\bm{0})&=\frac{1}{8\pi a^2}, {\rm \ \  and\ \ } G_{,yyyy}(\bm{0})=\frac{3}{8\pi a^2}.
\end{align}
We thus obtain
\begin{align}
M_{11}&=M_{22}=\frac{3}{8\pi a^2},\\
M_{12}&=M_{21}=\frac{\cos(2\theta_1-2\theta_2)-\cos(2\theta_1)-\cos(2\theta_2)}{4\pi R^2},
\end{align}
where we have switched back to $\theta_2=\pi-\theta'_2$. Note that $M_{12}$ is symmetric upon the exchange of $\theta_1$ and $\theta_2$ as it should be. Using eqn~(\ref{Hbyparts}), we obtain exactly
\begin{align}
H=4\pi\kappa a^2\frac{
4a^2C_\mathrm{r1}C_\mathrm{r2}f(\theta_1,\theta_2)
+3R^2(C_\mathrm{r1}^2+C_\mathrm{r2}^2)
}{9R^2+4f(\theta_1,\theta_2)a^4/R^2},
\end{align}
where $f(\theta_1,\theta_2)=\cos(2\theta_1)+\cos(2\theta_2)-\cos(2\theta_1-2\theta_2)$.
However, since the rods we are actually modeling are not point-like, the above formula is not meaningful for $R\simeq a$. We therefore take the leading order of $H$ in $1/R$, which describes the interaction quantitatively well for $R$ larger than a few times $a$. Replacing $a$ by $r_\mathrm{rod}$ yields the interaction $\tilde H_\mathrm{int}^{(0)}(R)$ given by eqn~(\ref{eq:Hint}).
Note that this result is symmetric upon exchanging $\theta_1$ and $\theta_2$ and invariant in changing $\theta_1$ into $\theta_1+\pi$ or $\theta_2$ into $\theta_2+\pi$, as it should be.

\subsection{Finite tension case}\label{tensioncase}

In the presence of membrane tension, for $\gamma\ne0$, the formalism in the previous section is still valid, provided the full Green function is used, i.e., the Green function of $\mathcal{L}=\nabla^2(\nabla^2-\xi^{-2})$ defined by eqn~(\ref{eq:Green}). In particular, eqns~(\ref{eq:M11})--(\ref{Hbyparts}) still hold.
The Green function in the presence of tension is given by~\cite{Weitz13}
\begin{equation}\label{eq:tensionGreen}
G(r)=-\frac{\xi^2}{2\pi}\left[K_0\!\left(\frac{r}{\xi}\right)+\ln r\right],
\end{equation}
where $K_0$ and $I_0$ (used below) are  modified Bessel functions.
Indeed, the most general rotationally symmetric solution of $\mathcal{L}G(\bm{r})=0$ is $G(r)=A_1+A_2\ln r+A_3 K_0(r/\xi)+A_4 I_0(r/\xi)$. The finiteness of $G$ at the origin requires $A_2=A_3$ and integrating eqn~(\ref{eq:Green}) around $\bm{r}=0$ yields $A_3=-\xi^2/(2\pi)$. Discarding the constant term $A_1$ and the term $A_4I_0(r/\xi)$ that diverges at infinity, we obtain eqn~(\ref{eq:tensionGreen}). Note that eqn~(\ref{eq:tensionGreen}) reduces to eqn~(\ref{eq:freeGreen}) for $r\ll\xi$.

Regularizing the fourth derivatives of the Green function as previously, we obtain
\begin{align}
G_{,xxxx}(\bm{0})&=\int_0^{2/a}\!\!\frac{q\,dq}{(2\pi)^2}\int_0^{2\pi}\!\!d\theta
\frac{q^4\cos^4\theta}{q^4+\xi^{-2}q^2}
\nonumber\\
&=\frac{3}{8\pi a^2}
-\frac{3}{32\pi\xi^2}\ln\left(
1+4\xi^2/a^2
\right).
\end{align}
Similarly, 
\begin{align}
G_{,xxxy}(\bm{0})&= G_{,xyyy}(\bm{0}) =0,\\ \nonumber
G_{,xxyy}(\bm{0})&=\frac{1}{8\pi a^2}
-\frac{1}{32\pi\xi^2}\ln\left(
1+4\xi^2/a^2
\right)
\nonumber\\
G_{,yyyy}(\bm{0})&=G_{,xxxx}(\bm{0}).
\end{align}
Using eqns~(\ref{eq:M11})--(\ref{Hbyparts}) with these elements, we obtain the graphs of $H_\mathrm{int}(R)$ shown in Sec.~\ref{sec:theory_tension} and the corresponding asymptotic behaviors $H_\mathrm{int}^{(1)}(R)$ and $H_\mathrm{int}^{(2)}(R)$.

\bigskip
\section{Interactions between parallel straight rod-assemblies}\label{sec:apthe2}

\begin{figure}
\centerline{\includegraphics[width=.9\columnwidth]{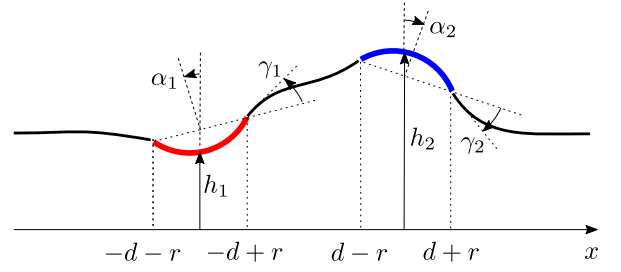}}
\caption{Geometrical parameters for the calculation of the interaction between two rod assemblies (side view).}
\label{fig:assemblies}
\end{figure}

We consider two parallel rod assemblies, one made of rods of curvature $C_\mathrm{r1}$ and the other one made of rods of curvature $C_\mathrm{r2}$. Recall that the rods are curved in the direction perpendicular to the axes of the rod assemblies. Let us define $r$, $\gamma_1$ and $\gamma_2$ in such a way that the curvatures $C_\mathrm{r1}$ and $C_\mathrm{r2}$ corresponds to angular variations $2\gamma_1$ and $2\gamma_2$ over the distance $2r=r_\mathrm{rod}$, respectively (see Fig.~\ref{fig:assemblies}). For the sake of simplicity, we regard the rod assemblies as infinite and homogeneous, i.e., we neglect extremity effects and the discrete character of the rods.

We fix the distance $R=2d$ between the centers of the two rod assemblies and we proceed to calculate the deformation energy stored in the membrane (taking into account membrane tension). We have, in principle, four degrees of freedom: the heights $h_1$ and $h_2$ of the assemblies with respect to the reference plane that is parallel to the membrane at infinity, and the tilt angle $\alpha_1$ and $\alpha_2$ of the normal to the assemblies relative to the normal to this reference plane. If we fix these four variables we have the following eight boundary conditions:
\begin{align}
\label{BC1}
\begin{tabular}{ll}
$h'_{1-}=\alpha_1-\gamma_1$,\quad~~~
&
$h'_{2-}=\alpha_2-\gamma_2$,
\\
$h'_{1+}=\alpha_1+\gamma_1$,
&
$h'_{2+}=\alpha_2+\gamma_2$,
\\
$h_{1-}=h_1-\alpha_1r$,
&
$h_{2-}=h_2-\alpha_2r$,
\\
$h_{1+}=h_1+\alpha_1r$,
&
$h_{2+}=h_2+\alpha_2r$.
\end{tabular}
\end{align}
Here the prime indicates differentiation with respect to $x$, the subscript `$1\pm$' refers to the position $x=-d\pm r$ and the subscript `$2\pm$' refers to the position $x=d\pm r$. The total energy of the membrane, per unit length and in the limit of small deformations, is given by the Helfrich Hamiltonian in the Gaussian approximation~\cite{Helfrich73}:
\begin{align}
\mathcal{H}=\int_M\!dx\left[\frac{\kappa}{2}h''^2(x)+\frac{\gamma}{2}h'^2(x)\right],
\end{align}
where $M=]-\infty,-d-r]\cup[-d+r,d-r]\cup[d+r,\infty[$. It 
has to be minimized with respect to the shape $h(x)$ of the membrane in $M$, but also with respect to $h_1$, $h_2$, $\alpha_1$ and $\alpha_2$, which corresponds to requiring that no forces nor torques act on the rod assemblies at equilibrium.

The corresponding conditions can be obtained directly by performing two integration by parts on the first variation of $\mathcal{H}$. We obtain
\begin{align}
\delta\mathcal{H}&=\int_M\!\left[\kappa h''''(x)-\gamma h''(x)\right]\delta h(x)\,dx
\nonumber\\
&-\Gamma_1\,\delta\alpha_1-\Gamma_2\,\delta\alpha_2
-F_1\,\delta h_1-F_2\,\delta h_2,
\end{align}
where we identify $\Gamma_1$, $\Gamma_2$, $F_1$ and $F_2$ as the torques and forces (per unit length) acting on the rod assemblies, respectively.  Requiring them to vanish we obtain a new set of eight boundary conditions, the conditions that must be satisfied at equilibrium:
\begin{align}
&\Gamma_1\equiv\kappa\left(h''_{1+}-h''_{1-}\right)
-\kappa r\left(h'''_{1+}+h'''_{1-}\right)
\nonumber\\&\qquad
+\gamma r\left(h'_{1-}+h'_{1+}\right)=0,
\\
&\Gamma_2\equiv\kappa\left(h''_{2+}-h''_{2-}\right)
-\kappa r\left(h'''_{2+}+h'''_{2-}\right)
\nonumber\\&\qquad
+\gamma r\left(h'_{2-}+h'_{2+}\right)=0,
\\
&F_1\equiv\kappa\left(h'''_{1-}-h'''_{1+}\right)+2\gamma\gamma_1=0,
\\
&F_2\equiv\kappa\left(h'''_{2-}-h'''_{2+}\right)+2\gamma\gamma_2=0,
\\
&h'_{1+}-h'_{1-}-2\gamma_1=0,
\\
&h'_{2+}-h'_{2-}-2\gamma_2=0,
\\
&h_{1+}-h_{1-}-r\left(h'_{1-}+h'_{1+}\right)=0
\\
&h_{2+}-h_{2-}-r\left(h'_{2-}+h'_{2+}\right)=0.
\end{align}
As for the membrane, it must satisfy the Euler-Lagrange equation
\begin{align}
h''''(x)-\xi^{-2} h''(x)=0.
\end{align}

Solving this linear, one-dimensional problem, yields the interaction energy per unit length of the rod assemblies
\begin{align}
f_\mathrm{int}(d)=2\sqrt{\kappa\gamma}\,\gamma_1\gamma_2\,e^{-2(d-r)/\xi},
\end{align}
which, after setting $2r=r_\mathrm{rod}$, $2d=R$, $2\gamma_1=C_\mathrm{r1}r_\mathrm{rod}$, $2\gamma_2=C_\mathrm{r2}r_\mathrm{rod}$ and multiplying by the length of the rod assemblies, yields eqn~(\ref{intbars}).

\end{appendix}


\end{document}